\documentclass{iaus}
\usepackage{graphicx}

\title[IR environments of masers] 
{The infrared environments of masers associated with star formation}

\author[De Buizer]   
{James De Buizer}

\affiliation{Gemini Observatory, Casilla 603, La Serena, Chile
\break email: jdebuizer@gemini.edu\\ 
}

\pubyear{2007}
\volume{242}  
\pagerange{1--8}
\date{?? and in revised form ??}
 \jname{Astrophysical Masers and Their
Environments} \editors{J. Chapman \& W. Baan, eds.}
\begin{document}

\maketitle

\begin{abstract}
The near infrared (1-2~$\mu$m) and the thermal infrared
(3-25~$\mu$m) trace many of the environments in which masers are
thought to reside, including shocks, outflows, accretion disks, and
the dense medium near protostars. After a number of recent surveys
it has been found that there is a higher detection rate of mid-IR
emission towards masers than cm radio continuum emission from UC~HII
regions, and that the mid-IR emission is actually more closely
cospatial to the maser locations. A high percentage of water and
methanol masers that are not coincident with the UC~HII regions in
massive star forming regions are likely to be tracing outflows and
extremely young high mass stars before the onset of the UC~HII
region phase. After a decade of groundwork supporting the hypothesis
that linarly distributed class II methanol masers may generally
trace accretion disks around young massive stars, compelling
evidence is mounting that these masers may generally be associated
with outflows instead. Substantiation of this claim comes from
recent outflow surveys and high angular resolution mid-IR imaging of
the maser environments.

\end{abstract}


In the proceedings of the last maser meeting in Brazil in 2001, I
discussed the largely untapped potential of the infrared spectral
regime as a tool for understanding the nature of the circumstellar
environments of masers associated with star formation (De Buizer
2002a). In that article, I described the early stages of the work I
had started using the infrared (IR) to study maser environments, and
gave several possible paths for future work in the field. In the
intervening years since that meeting and this one, I have explored
several of those paths in detail, and will present in this article
some of the highlights of that work.

\section{The close association of masers and IR emission}

In the Brazil proceedings article I described how the near infrared
(1-2~$\mu$m) and the thermal infrared (3-25~$\mu$m) trace everything
from the direct photospheric emission from young stellar sources to
the relatively cool dusty environments that can be distributed far
from their parent stars. The thermal infrared in fact has been shown
to trace dusty circumstellar environments from temperatures of
$\sim$1500~K at 3~$\mu$m, which is very close to a star and near the
dust sublimation temperature, to cool dust 10s of thousands of AU
away at wavelengths near 25~$\mu$m. Because of this, it was
suggested that the IR would be a good tracer of the emission from
many of the environments in which masers were thought to reside:
shocks fronts, outflow shocks, and accretion disks.

But how well does the infrared trace the environments of maser
emission? There now exist a number of studies at several different
wavelengths to help address this question. Since it is easiest to
obtain radio continuum observations of maser regions, the most
common studies of maser environments have traditionally involved
searches for ionized gas emission in the form of UC~HII regions and
partially ionized outflows. Well before the Brazil meeting it was
known that the percentage of maser sources directly associated with
cm radio continuum emission was not very high, which was remarkable
for a phenomenon that is suppose to trace young massive star
formation.

From studies with reasonable detection limits at a variety of
wavelengths, one can piece together the dominant wavelengths at
which maser environments can be seen and studied. As mentioned
above, the cm radio continuum detection rate towards methanol and
water masers is low, and has been found to be about 20\% from the
recent surveys of Walsh et al. (1998) and Beuther et al. (2002).
Whereas in the surveys of De Buizer et al. (2000), Walsh et al.
(2001), and De Buizer et al. (2005) it was found that the detection
rate of mid-IR (typically 10 and 18~$\mu$m) emission towards class
II methanol masers is $\sim$70\%, and towards water masers
$\sim$80\%. Coarser spatial resolution observations in the sub-mm
and mm are showing detection rates approaching 100\% towards both
water and methanol masers (Walsh et al. 2003; Beuther et al. 2002).
Hence, all masers appear to be more closely associated with regions
of hot to warm (300-30K) thermal dust emission than ionized gas
emission.

Furthermore, De Buizer et al. (2005) showed that mid-IR emission
when detected in a maser region was more closely associated with
actual maser locations than the radio continuum emission detected
towards those same maser regions (i.e., Figure 1b). The survey of
Hofner \& Churchwell (1996) showed that the median separation
between water maser and the radio UC HII regions they detected is
$\sim$18800~AU. The median separation between water masers and
mid-IR sources is $\sim$8700~AU (De Buizer et al. 2005). Therefore,
not only is there a higher detection rate of mid-IR emission towards
masers than cm radio continuum emission, mid-IR emission is actually
more closely cospatial to the maser locations.

\section{Using masers to pin-point extremely young massive stars}

A large subset of water and class II methanol masers are found in
regions of massive star formation, but are not coincident with the
UC~HII regions found there. What are these masers tracing?

In the case of the G29.96-0.02 region, the methanol and water masers
that are offset from the cometary UC~HII region and were instead
found to be coincident with a hot molecular core discovered in the
ammonia line images by Cesaroni et al. (1994). Many hot molecular
cores are high mass protostellar objects (HMPOs) which are an
extremely early stage of massive stellar birth. An HMPO consists of
a massive protostar surrounded by a thick envelope of accreting dust
and gas. They are compact sources seen in radio-wavelength ammonia
(or molecular line) images but are so young that they have not had
time to ionize their surroundings, and hence typically have little
to no detectable radio continuum emission of their own. Other cases
similar to G29.96-0.02, where masers were found associated with HMCs
and offset from UC~HII regions, lead to the idea that some masers
species may generally trace these very young sources at the earliest
stages of massive star formation.

Some of the first models of the HMPO phase of massive star formation
that were constructed (e.g. Osorio et al. 1999) showed that by
changing parameters such as spectral type, accretion rates, and core
radius, one can construct a variety of model spectral energy
distributions (SEDs) for HMPOs. Throughout an array of physically
reasonable parameter space, the SEDs peak around 80~$\mu$m on
average. On the Rayliegh-Jeans side of the SED ($>$80~$\mu$m), there
is little change in the shape of the SED when parameters are
changed. However, on the Wien side of the SED ($<$80~$\mu$m), there
is considerable change in SED shape with the variation of each
parameter, especially if one could observe the mid-IR emission of
these sources (see Figure 1a and 1c). The most noticeable changes in
the HMPO SEDs are between 3 and 30~$\mu$m, and in the depth and
shape of the 10~$\mu$m silicate feature. This modeling showed that
the mid-infrared could be key to understanding the properties of the
youngest massive stars, if we could detect the mid-IR emission from
such embedded and distant objects.

\begin{figure}
 \includegraphics[width=5.3in,angle=0]{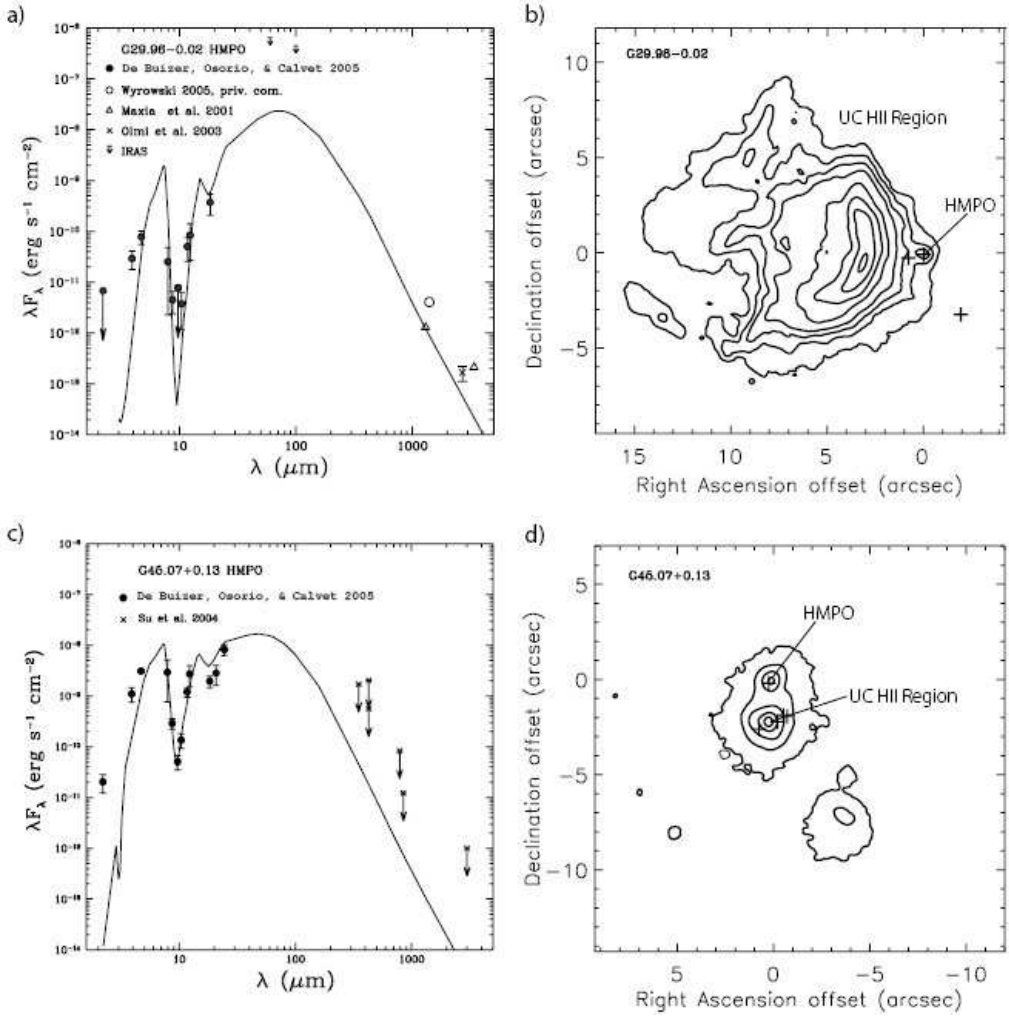}
  \caption{\textbf{(a)} Observed and model SEDs for the G29.96-0.02 HMPO. The
different symbols represent the observed values of the flux
densities from the references indicated. Error bars are 3-$\sigma$
for the IR data. Upper limits are represented by arrows. The solid
line represents the best fit model obtained with L$_{\star} =
1.8\times10^4$ L$_{\odot}$, the equivalent luminosity of a B0 star.
\textbf{(b)} A contour plot of the 11.7~$\mu$m image of the
G29.96-0.02 field with the extended mid-IR emission from the UC HII
region and compact mid-IR emission of the HMPO indicated. Plus signs
represent the location of water masers from Hofner \& Churchwell
(1996). The masers are not associated with the UC HII region, but
some are associated with the HMPO. \textbf{(c)} Same as in (a) but
for the G45.07+0.13 HMPO. The solid line represents the best fit
model obtained with L$_{\star} = 2.5\times10^4$ L$_{\odot}$.
\textbf{(d)} Same as in (b) but for G45.07+0.13. In this case, there
are some water masers associated with the UC HII region, however the
clump of masers to the north are associated with an
HMPO.}\label{fig:fig1}
\end{figure}

To see if we could indeed detect these HMPOs in the mid-IR and test
the hypothesis that masers offset from UC~HII regions may be
associated with HMPOs, I began my search in the mid-IR with the
prototypical source, G29.96-0.02. These observations led to the
first confirmed direct detection of a HMC at mid-IR wavelengths (De
Buizer et al. 2002a), proving that indeed some HMPOs are bright
enough to observe in the mid-IR. To further test the hypothesis that
non-radio continuum emitting sources of maser emission could be
associated with HMPOs, a mid-infrared study of several fields from
the survey of water maser and UC~HII regions of Hofner \& Churchwell
(1996) was performed in an attempt to find more mid-infrared bright
HMPOs. Concentrating on the fields that have UC~HII regions with
water masers well offset, this survey led to the detection of
mid-infrared emission from the locations of two HMPO candidates,
G11.94-0.62 and G45.07+0.13 (De Buizer et al. 2003). These
observations seemed to support the idea that in some cases masers
can indeed be used to find the locations of HMPOs.

At the Brazil meeting I claimed that these mid-IR-bright sources
held the promise that, if one could perhaps create a well-sampled
SEDs from observations of a HMPO at many wavelengths in the
mid-infrared, the data could then be fit with the new HMPO models to
derive accurate physical parameters, such as mass, luminosity and
accretion rate for these youngest massive stars.

This was attempted by De Buizer, Osorio, \& Clavet (2005), for the
three above mentioned sources: G29.96-0.02, G11.94-0.62, and
G45.07+0.13. Though the modeling performed on these sources led to
relatively accurate luminosities, degeneracies and a large number of
free parameters prevent the accurate assessment of physical
parameters from SED modeling alone and can only constrain such
parameters. More recent and more detailed SED modeling of YSOs by
Robitaille et al. (2007) have come to the same conclusion.

Interestingly, of the three sources modeled in De Buizer, Osorio, \&
Calvet (2005), G11.94-0.62 has a luminosity to small to be a true
HMPO. Though this source lies near a UC~HII region, is cospatial
with water maser emission, and has a deep silicate feature in the
mid-infrared indicative of a highly embedded star, it is likely a
young and embedded intermediate mass YSO. Hence, there exist
intermediate mass YSOs that have similar observational
characteristics to HMPOs that may contaminate HMPO studies.

From all of this, one can conclude that some masers offset from
UC~HII regions can indeed be associated with HMPOs. However, in De
Buizer et al. (2005) it is shown that some of these masers are
likely offset from known YSOs because they are associated with
outflow, and some, as just mentioned, may be associated with
less-massive embedded YSOs.

\section{Maser disks vs. maser outflows}

Some maser species, class II methanol in particular, have been
suggested to be associated with circumstellar disks around massive
young protostars. Norris et al. (1993) discovered that 45\% of
methanol masers groups tend to be distributed in the sky in linear
structures. Occasionally these linearly distributed masers have
apparent velocity gradients along the maser distributions that
suggest rotation. Observations by Walsh et al. (1998) confirmed that
methanol masers are linearly distributed in approximately half the
cases, but that velocity gradients were not a general feature.

However, in the article by Norris et al. (1993) it was suggested
that these linearly distributed methanol masers exist in, and
delineate, edge-on circumstellar disks around massive stars. In
Brazil I presented results from my low resolution (1.0--2.0'')
mid-IR imaging survey of star forming regions with methanol maser
emission, which contained 10 sites of linearly distributed masers.
Three of these maser locations were coincident with young stellar
objects with extended mid-IR emission elongated at the same position
angle as their methanol masers distributions (De Buizer et al.
2000). These sources were suggested to be dusty circumstellar disks,
thereby apparently adding credence to the disk hypothesis for
linearly distributed masers. However, one of these sources was later
observed at high angular resolution ($<$0.5'') with Keck and found
to be 3 individual mid-IR sources arranged in a linear fashion, and
not a disk (De Buizer et al. 2002b).

To further test the maser/disk hypothesis, several sites of linearly
distributed methanol masers were observed to search for outflows.
According to the standard model of accretion, during the phase of
stellar formation where the star is being fed from an accretion
disk, it is also undergoing mass loss through a bipolar outflow. The
bipolar outflows are perpendicular to the plane of the accretion
disk, and along the axis of rotation. Wide-field images of the sites
of linearly distributed methanol masers were obtained using the
2.12~$\mu$m H$_2$ ($\nu$=1–-0) S(1) line as the outflow diagnostic
(De Buizer 2003). Surprisingly, in 12 of 14 fields where H$_2$
emission was detected the emission was not aligned perpendicular to
the maser distributions as expected, but instead was parallel
(Figure 2). This seemed to suggests that the methanol masers
delineate outflows rather than circumstellar accretion disks.

\begin{figure}
 \includegraphics[width=5.25in,angle=0]{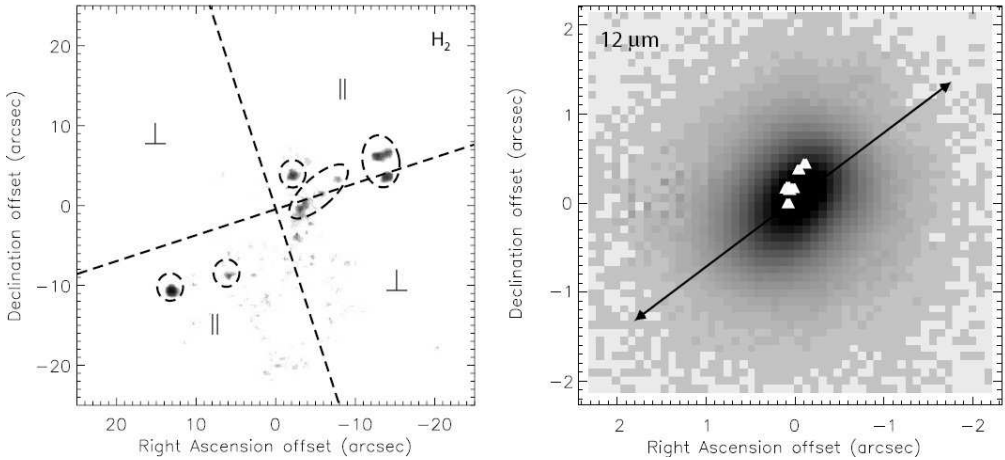}
  \caption{\textbf{(Left)} The H$_2$ line image from De Buizer
(2003) of the $\sim$50'' field of G318.95-0.20 centered on the
methanol maser location. The dashed straight lines intersect at the
maser location and show the parts of the field perpendicular to
($\bot$) and parallel to ($\|$) the methanol maser distribution
angle. Dashed ellipses show the location of real H$_2$ emission
(other faint signatures are due to poor subtraction of continuum
emission). The H$_2$ emission is distributed in the quadrants
parallel to the linear methanol maser distribution angle.
\textbf{(Right)} The 12 $\mu$m image of the $\sim$5'' field of
G318.95-0.20. The methanol masers of Norris et al. (1993) are shown
as white triangles. The mid-IR emission is elongated at the same
angle as the methanol maser distribution, however this is the
direction of the outflow seen in H$_2$. The arrow shows the
direction of the SiO outflow (De Buizer et al. in prep). The
similarity of the methanol maser distribution angle and mid-IR
emission elongation angle with the outflow angle suggest that both
the maser and the mid-IR emission are outflow related, and not
directly tracing a circumstellar disk.}\label{fig:fig2}
\end{figure}

The interpretation of the results from De Buizer (2003) remained
ambiguous because 2.12~$\mu$m H$_2$ line emission can be excited
both by outflow shocks and by UV excitation.  The overall morphology
of the H$_2$ emission seen the fields imaged by De Buizer (2003)
does not resemble the bipolar outflows seen around young, low-mass
stars. It was cautioned that perhaps some of the H$_2$ emission was
perhaps due to excitation from UV emission by other massive stars in
the star forming region. Without additional evidence, it could not
be said with certainty which mechanism is stimulating the H$_2$
emission near these high-mass protostars, nor definitively link the
outflows to the methanol masers. However, 86\% of the sources showed
H$_2$ emission parallel to their maser distribution angles, and the
probability of this occuring by chance is low. Nonetheless, there
was sufficient doubt to justify observing these sources in an
independent outflow indicator.

SiO is a good shock tracer because its abundance is enhanced by
factors of up to 10$^6$ behind strong shocks (i.e., Avery \& Chiao
1996). Using the JCMT to try to detect the presence of outflows in
the SiO (6-5) transition, nine sources from the H$_2$ survey of De
Buizer (2003) were observed and SiO emission was found in seven of
those fields (Feldman et al. 2005). Many of the stronger detections
have line profiles with a relatively narrow core on top of wide
wings which are characteristic of outflows. To follow this up, 5 of
these JCMT sources were recently observed with ATCA to map out the
outflows in the SiO (2-1) transition. Preliminary results from these
observations show that the SiO emission is in all cases distributed
at an angle close to that of the maser distribution angles and H$_2$
emission position angles for each source. In 4 of the 5 cases these
SiO maps show clear signs of a single outflow with red- and
blue-shifted lobes. In the fifth case the velocity structure is more
complex, indicating perhaps multiple outflows are present, however
the overall structure of the SiO is parallel to the maser position
angle (De Buizer et al. in prep). These SiO observations clearly
indicate that the H$_2$ emission observed in Buizer (2003) is indeed
outflow related. These observations also lend convincing support to
the idea that the methanol masers in linear distributions are
directly associated in some way with the outflow from their parent
stars.

Recent high angular resolution observations of individual young
massive stellar objects with masers (De Buizer 2006, De Buizer 2007,
and as yet unpublished data) have revealed that many sources that
are elongated in their mid-IR emission are not disks, but instead
are elongated in the direction of outflow. The best example of this
is G35.20-0.74 (De Buizer 2006), which shows an outflow cavity in
mid-IR continuum emission unmistakeably similar to the outflow jet
seen at various other wavelengths. It is believed that sources such
as these are too embedded to directly detect their accretion disks
even at mid-IR wavelengths. However if the outflows clear away
enough material in the natal cloud surrounding a massive young
stellar object, we are then able to see these cavities and the
mid-IR emission of the warm cavity walls. More developed bipolar
cavities should exist where the material above and below the
accretion disk is well-cleared as the opening angles of the outflow
cavities widen. In this case the definition of what is the surface
of a outflow cavity and what is the surface of a flared accretion
disk may become blurred, and the source could appear as a mid-IR
silhouette disk. These disks would be similar to the near-IR
silhouette disks seen in Orion (i.e., McCaughrean et al. 1998),
however the accretion disks would still be so optically thick at
their mid-planes that one could not detect the direct emission from
the disk itself. Detections of such sources may have already been
made; possible examples of such sources are the massive young
stellar object in M17 (Chini et al. 2004) and IRAS 20126+4104 (De
Buizer 2007).

In the case of G35.20-0.74 the OH, water, and methanol masers appear
to delineate the edges of these outflow cavity walls as seen in the
mid-IR. Therefore, it is a possibility that the linear arrangements
of water and methanol masers associated with other sources are
tracing cavity walls where there are oblique shocks for the
collisional pumping of water masers and a sufficient mid-IR thermal
bath of photons to radiatively pump the methanol masers.

Observations in the mid-IR of massive young stellar objects that
only show signs of emission from their outflow cavities may also
help interpret the mid-IR emission from more embedded HMPOs. Some
HMPOs are not mid-IR bright, while others are. In many cases this
may simply be a temperature issue, i.e. that some HMPOs are too cold
to be seen at mid-IR wavelengths. However, HMPOs that are detectable
in the mid-IR may just be sources where the outflows from the
central stars are more or less pointed toward us. In this way, we
are seeing through the cavity and closer to the central heating
source since the outflow is clearing out material along our line of
sight. Therefore the mid-IR emission observed to be coming from
HMPOs like G29.96-0.02 may be tracing beamed emission in outflow
cavities and not reprocessed emission coming from the entire core,
as is seen at longer wavelengths. If this is true, SED models of
such sources would be influenced by this and could lead to poor fits
unless accounted for.

\section{Conclusions}

A lot has been learned about the infrared environments of masers
since the 2001 Brazil maser meeting. The IR regime has been shown to
complement radio continuum observations and offer new insights into
maser environments.

Methanol, hydroxyl, and water masers appear to be most closely
associated with regions of hot to warm (300-30 K) thermal dust
emission traced by mid-IR and sub-mm emission. As such, studies of
the IR environments of masers will continue to offer much insight
into the relationship between masers and the massive star formation
process, {\it especially at high ($<$0.5'') angular resolution}
where one can hope to separate IR emission from disks, envelopes,
outflows, and other nearby stellar sources.

It has been found that water and methanol masers offset from UC HII
regions trace mid-IR sources that are in some cases HMPOs. Therefore
masers that appear to be isolated from traditional massive star
formation tracers such as radio continuum emission may be great
locations to find more HMPOs. If enough observations are made over
the full SED of HMPOs, physical properties of these sources can be
derived from accretion models. However, such SED modeling alone can
only place constraints and limits on most of these physical
parameters.

Evidence is building that shows that linear distributions of
methanol masers are not generally tracing disks around young massive
stars. These maser distributions were found to be dominantly at the
same angle as extended near-IR H$_2$ emission, which is a potential
outflow indicator. Follow-up observations in the mm using SiO as an
outflow tracer appear to confirm the notion that outflows are
dominantly at the same position angles as the linear methanol maser
distributions of massive young stellar objects. This further implies
a {\it general, direct physical relationship between the methanol
masers and outflows}, rather than disks, however these observations
do not rule out the possibility that masers may trace circumstellar
disks or other phenomena in some cases.

Finally, high-resolution imaging is showing circumstellar mid-IR
continuum emission commonly comes from outflows, not disks. This
mid-IR emission is often found to be co-spatial with masers, and
therefore further suggests that there is a general tendency for
masers, especially water and methanol, to be often directly
associated with outflows.


\end{document}